# Implementing OpenSHMEM for the Adapteva Epiphany RISC Array Processor

James A. Ross[1] and David A. Richie[2]
[1]U.S. Army Research Laboratory, Aberdeen Proving Ground, MD
[2]Brown Deer Technology, Forest Hill, MD
james.a.ross176.civ@mail.mil, drichie@browndeertechnology.com

**Abstract**
The energy-efficient Adapteva Epiphany architecture exhibits massive many-core scalability in a physically compact 2D array of RISC cores with a fast network-on-chip (NoC). With fully divergent cores capable of MIMD execution, the physical topology and memory-mapped capabilities of the core and network translate well to partitioned global address space (PGAS) parallel programming models. Following an investigation into the use of two-sided communication using threaded MPI, one-sided communication using SHMEM is being explored. Here we present work in progress on the development of an OpenSHMEM 1.2 implementation for the Epiphany architecture.

*Keywords:* SHMEM, PGAS, Adapteva Epiphany, many-core, energy-efficient, one-sided communication, NoC

## 1 Introduction

The Adapteva Epiphany MIMD architecture [1], currently realized in the inexpensive Parallella platform, remains a challenge to program. Contributing factors include the limited core memory (32 KB shared instruction and data), low off-chip bandwidth, an unfamiliar proprietary software stack, and current inability to access host system calls within the asymmetric hybrid platform architecture. A general overview of the Epiphany architecture appears in Figure 1. The vendor-developed multi-core e-lib interface within Epiphany SDK (eSDK) requires rewriting parallel applications in order to take advantage of the underlying hardware features such as the on-chip dual-channel DMA engines and 2D NoC topology. However, such applications cannot be reused on other platforms. The e-lib interface also lacks most of the multi-core primitives found in the OpenSHMEM interface so few direct comparisons are available.

Computer architectures like Epiphany achieve excellent computational energy efficiency. The use of SRAM instead of off-chip DRAM decreases power consumption and reduces memory latency, while addressing the memory wall problem found in symmetric multiprocessor (SMP) architectures. The bandwidth scales with the number of cores in the same manner that DRAM bandwidth scales with the number of sockets on a distributed CPU cluster. The architecture may be tiled so that a larger coprocessor can be created by placing additional processors on a circuit board and connected without additional glue logic. Conceptually, the greatest challenges for effectively using the Epiphany cores are from the limited SRAM as well as the efficient execution of inter-processor communication primitives.

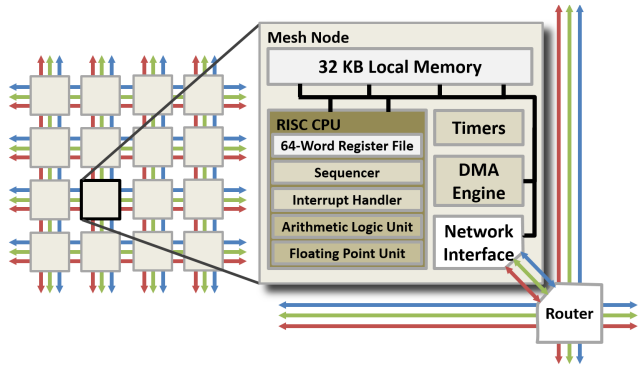

**Figure 1:** The Epiphany-III coprocessor includes 16 processing elements connected across a 2D network-on-chip. Each core contains 32 KB local memory and a dual-channel DMA engine.

In previous work, we demonstrated the use of a threaded MPI implementation to achieve high performance using a standard parallel programming API for the Epiphany architecture [2], [3]. The OpenSHMEM 1.2 standard provides excellent one-sided communication routines well-suited for Epiphany when executed in a SPMD manner. For the Epiphany architecture, the OpenSHMEM API provides improved data referencing semantics and reduced interface complexity compared to MPI, thus reducing code size and increasing application performance. We present here the status of the implementation of the ARL OpenSHMEM implementation for Epiphany. Most of the OpenSHMEM routines have been implemented, including all of the accessibility functions, atomics (add and fetch-and-operate), wait operations, reductions, locks, block data copy, and elemental data copy routines.

## 2 ARL OpenSHMEM Implementation for Epiphany

The eSDK uses a 2D identifier for the row and column of the core within the chip. This API choice restricts applications to use rectangular sections of the chip rather than arbitrary or odd work group sizes. There is no abstraction between a virtual process identifier and the physical domain and this is problematic for future architectures where there may be broken or disabled cores within a larger 2D array. With the OpenSHMEM API, the one-dimensional virtual computational topology abstracts away the physical location and memory address. Calculating the physical address of cores within a workgroup is comprised of trivial logical and integer operations. Direct comparisons between APIs is challenging because device code within the eSDK library contains only a subset of the routines used in OpenSHMEM: remote address calculation, a global barrier, block data memory copying, and multi-core locks.

Compared to the previous work on MPI for the architecture, the simplicity of the explicit type-specialization of OpenSHMEM routines enables more compact implementation, saving limited core memory resources. Additionally, one-sided communication and weaker synchronization requirements reduce the effective code size of an application compared with the use of explicit two-sided threaded MPI routines. There are also fewer specialized routines in OpenSHMEM to build out a full implementation. The MPI specification makes no assumptions for symmetric memory allocation, so additional inter-processor coordination may be required for correct remote address calculation.

Symmetric memory management is one of the most challenging aspects of the standard for the Epiphany architecture because there is no translation between logical and physical address of local memory and there is no tracking. As a consequence, calculating remote addresses is trivial and does not require inter-core coordination. The assumptions of symmetric allocation within SHMEM lead to less code and inter-processor coordination compared to the MPI implementation. SHMEM memory management routines are presently implemented using UNIX brk/sbrk for linear ordered allocation and imposes rules on the ordering of reallocation and freeing. We will address this limitation in the future to allow allocation accounting consistent with a conventional malloc.

For the fully collective barrier, the Epiphany hardware wait-on-AND (WAND) barrier and interrupt service routine are used. This is much higher performance than a software barrier, but has limited pragmatic usefulness when not all cores are being used. Additionally, for a proposed 1000+ core processor it will be expected that some disabled cores must be removed in software from the topology, making the development of an efficient scalable software barrier critical. The implementation of more efficient collective and strided barriers are currently under development. The WAND barrier and linear software barrier from the eSDK complete in 0.1 and 2.0 microseconds, respectively (20x speedup).

We developed an optimized block data copying method for fast inter-core data transfer. It utilizes the Epiphany hardware loop feature and unrolled double-word loads and stores in order to achieve higher bandwidth and lower latency than the eSDK. The eSDK utilizes the DMA engine and then spins on the DMA status register to wait for completion. Spinning on the status register causes a degradation in performance. The DMA engine has significant setup cost so small transfers are costly and hardware prevents achieving peak performance. The ARL OpenSHMEM block data copying methods achieve 2.1-9.9x speedup for all transfer sizes and data types compared to using the synchronous DMA copy method in the eSDK.

Multi-core locks within both the eSDK and ARL OpenSHMEM both utilize the test-and-set instruction so there is little performance difference. The eSDK does not provide any interfaces to atomic operations or reductions. Few applications have yet been developed with the ARL OpenSHMEM interface, however, we have written an optimized dot product application using the shmem_float_sum_to_all reduction method and measured the 16-core reduction to take 0.632 microseconds for the largest vector that could fit within core memory. This enabled the processor to achieve a collective 16.8 GFLOPS (67.2 GB/s) or 87% of peak performance. We believe this is a very positive result demonstrating the architecture is capable of achieving high efficiency with both computation and multi-core communication for applications that are challenging to bandwidth-constrained SMP architectures.

# 3   Current Development

Similar to the threaded MPI programming model for Epiphany, the SHMEM implementation will be executed as a threaded coprocessor task and is supported by the COPRTHR software stack targeting Epiphany. However, recent work in the development of COPRTHR-2 has demonstrated the ability to execute threaded applications directly without the co-development of a host application. A re-designed C compilation model and runtime includes support for the transparent indirect execution of host system calls, effectively exposing host Linux services to Epiphany cores. This software stack will enable SHMEM benchmarks to execute unmodified. Early experiments show that most of the OpenSHMEM C examples can be executed on Epiphany with absolutely no software changes already. This has the effect of significantly reducing the software development time for the Epiphany processor because portable SHMEM code can be executed without explicit host code. In addition, COPRTHR-2 memory management functionality will replace the UNIX brk/sbrk calls currently used for SHMEM memory management, thus eliminating the restrictions on operation ordering. Applications, specific subroutine performance results, and additional implementation details will be presented shortly.

It is our intention that ARL OpenSHMEM for Epiphany will be released as open source software through GitHub on the U.S. Army Research Laboratory account [4] with the hope that it will be useful to the Parallella community.